\documentclass[conference,letterpaper]{IEEEtran}
\usepackage{graphicx,color}
\usepackage{cite}
\usepackage{setspace} 
\usepackage{amsmath}
\usepackage{amsmath,amsthm,amssymb}
\usepackage{multirow}
\usepackage{hhline}
\usepackage{epsfig}
\usepackage{subcaption}
\usepackage{epstopdf}
\usepackage{verbatim}
\usepackage{algorithm}
\usepackage{algorithmicx}
\usepackage{algpseudocode}
\usepackage{etoolbox}
\usepackage{cases}
\usepackage{csquotes}
\usepackage{mathtools,stmaryrd}
\SetSymbolFont{stmry}{bold}{U}{stmry}{m}{n}
\usepackage{bbm}
\usepackage{lettrine}
\usepackage{array}
\usepackage{booktabs}

\makeatletter
\newcommand*{\algrule}[1][\algorithmicindent]{%
  \makebox[#1][l]{%
    \hspace*{.2em}
    \vrule height .75\baselineskip depth .25\baselineskip
  }
}

\newcount\ALG@printindent@tempcnta
\def\ALG@printindent{%
    \ifnum \theALG@nested>0
    \ifx\ALG@text\ALG@x@notext
    \else
    \unskip
    \ALG@printindent@tempcnta=1
    \loop
    \algrule[\csname ALG@ind@\the\ALG@printindent@tempcnta\endcsname]%
    \advance \ALG@printindent@tempcnta 1
    \ifnum \ALG@printindent@tempcnta<\numexpr\theALG@nested+1\relax
    \repeat
    \fi
    \fi
}
\patchcmd{\ALG@doentity}{\noindent\hskip\ALG@tlm}{\ALG@printindent}{}{\errmessage{failed to patch}}
\patchcmd{\ALG@doentity}{\item[]\nointerlineskip}{}{}{} 
\makeatother

\allowdisplaybreaks

\begin{document}

\title{Energy-efficient UAV movement and\\ user-UAV association in multi-UAV networks}

\author{\IEEEauthorblockN{Subhadip Ghosh, Priyadarshi Mukherjee, and Sasthi C. Ghosh}

\IEEEauthorblockA{Advanced Computing
 and Microelectronics Unit\\ Indian Statistical Institute, Kolkata 700108, India\\
Emails: cs2327@isical.ac.in, priyadarshi@ieee.org, sasthi@isical.ac.in}}

\maketitle
\begin{abstract}
These days, unmanned aerial vehicle (UAV)-based millimeter wave (mmWave) communication systems have drawn a lot of attention due to the increasing demand for faster data rates. Given the susceptibility of mmWave signals to obstacles and high propagation loss of mmWaves, ensuring line-of-sight (LoS) connectivity is critical for maintaining robust and efficient communication. Furthermore, UAVs have limited power resource and limited capacity in terms of number of users it can serve. Most significantly different users have different delay requirements and they keep moving while interacting with the UAVs. In this paper, first, we have provided an efficient solution for the optimal movement of the UAVs, by taking into account the energy efficiency of the UAVs as well as the mobility and delay priority of the users. Next, we have proposed a greedy solution for the optimal user-UAV  assignment.
After that, the numerical results show how well the suggested solution performs in comparison to the current benchmarks in terms of delay suffered by the users, number of unserved users, and energy efficiency of the UAVs.
\end{abstract}

\vspace{-2mm}

\IEEEpeerreviewmaketitle

\section{Introduction}
In today's times, unmanned aerial vehicles (UAVs) appear to be a very promising research direction due to their multitude of applications in fields such as military, disaster management, and communications \cite{applications}. In this context, especially the low-altitude UAVs offer several advantages such as easy deployment, cost-effectiveness, and lightweight design. However, one major disadvantage of UAVs is that they are battery-operated, i.e., they have limited energy resources. In recent years, millimeter wave (mmWave) communications have also gained attention because of their higher carrier frequencies and better transmission characteristics. However, due to shorter wave length, it is susceptible to being easily blocked by obstacles and therefore has a lower coverage area \cite{mmw}. Thereafter, consideration of rotary-wing UAVs as flying base stations can help in mitigating these challenges of mmWave communication \cite{zeng2019energy}.

In UAV assisted mmWave communications systems, the key goal is to determine the optimal movement of the UAVs and establish the efficient user-UAV links. The work in \cite{k-means-ref} implements the K-means algorithm by using distance as metric for user-UAV assignment. The authors in \cite{li2022geometric} propose a modified K-means algorithm by using path-loss as the metric. Note that, both \cite{k-means-ref} and \cite{li2022geometric} assume that the users are static and that an UAV can serve any number of users. However, in practical scenario there is always a limitation on serving capacity of the UAVs and users are often mobile and not static. The work in \cite{li2022geometric} considers both LoS and NLoS links, but due to higher path-loss in mmWave communications, the NLoS links are almost infeasible  \cite{sau2024drams}. Also, to avoid interference among the co-existing users, the use of orthogonal frequency division multiple access (OFDMA)  is a very effective and practical approach for subcarrier assignment
\cite{wang2024adaptiveOFDMA}, \cite{wang2018spatialOFDMA}, where the number of subcarrier frequencies is limited. Because of this limitation of OFDMA, distributing the users over all UAVs becomes a real challenge. Accordingly, the work in \cite{shen2024flexible} considers this aspect to propose  efficient resource allocation schemes to enhance the overall system performance. The work in \cite{malinen2014balanced} propose a balanced K-means clustering method by using the Hungarian algorithm \cite{kuhn1955hungarian} and the authors in \cite{kuila2012energy-LB_clustering} propose a greedy approach for balancing the cluster sizes. But all the works assume that any user can be assigned to any cluster, i.e., any user is serviceable by all the UAVs. But, because of the infeasibility of NLoS in mmWave-based scenario, that may not be the case. Moreover, \cite{li2022geometric} and \cite{ma2024joint} propose user-UAV assignment solely based on their respective throughput performances. But, due to limited capacity of the UAVs and NLoS scenario, these approaches can result in multiple users not getting served and also higher imbalance of cluster sizes. The work in \cite{zeng2020resource} proposed UAV trajectory planning and resource allocation schemes considering the delay requirements of the users without considering the capacity of the UAVs. \cite{yin2023joint} proposed resource allocation and user-UAV assignment considering user delay taking distance as the metric. However none of them has taken the user mobility into consideration.

Hence, by considering all these aspects of an UAV assisted mmWave-based system, in this work, for a multi-UAV multi-user network, we propose computationally efficient algorithms for obtaining balanced clusters, energy-efficient movement of UAVs, and capacity-constrained user-UAV assignment by taking into account both the delay priority and mobilty of the users. Specifically, we propose a priority aware K-means based algorithm for the initial clustering while maximizing throughput, capturing user mobility, and also, keeping the cluster size below a pre-defined limit. Moreover, we take into account the aspect of energy-efficiency, while considering the movement of the UAVs. Furthermore, we also propose a greedy capacity-constrained algorithm for user-UAV assignment while considering both delay priority and mobility of users with taking care of throughput maximization, without repositioning the UAVs. Finally, we discuss the computational complexity of the proposed solution and the numerical results support the superiority of the same in terms of delay suffered by the users, number of unserved users, and energy efficiency of the UAVs in comparison to BT-kmeans \cite{li2022geometric} and balanced K-means \cite{malinen2014balanced} algorithms.

This paper is organized as follows. In section \ref{sys_model}, we have discussed the system model and problem formulation. In section \ref{prop_str}, we present the proposed UAV movement and user-UAV association algorithms. We discuss the simulation results and compare them with the existing benchmarks in section \ref{result}. Finally, concluding remarks are given in section \ref{con}.

\section{System model}\label{sys_model}
In this section, we discuss both the system model considered and the formulation of the problem.
\vspace{-1mm}
\subsection{Network Topology}

\begin{figure*}
    \centering
    \includegraphics[width=0.8\linewidth]{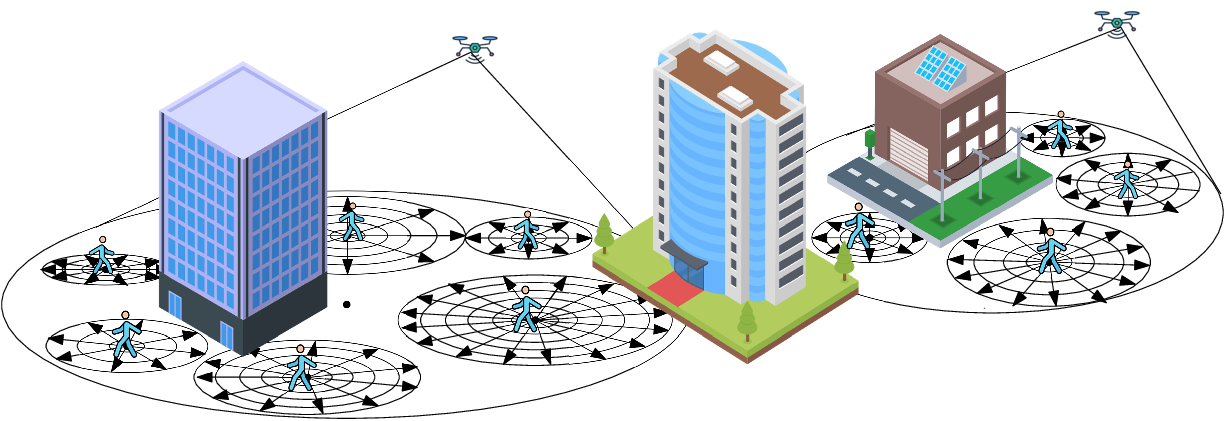} 
    \caption{Considered system model.}
    \label{smodel}
    \vspace{-5mm}
\end{figure*}

As can be seen in Fig. \ref{smodel}, here we consider a multi-user multi-UAV scenario. Specifically, we have a system of $M$ UAVs and $K$ users, where $\mathcal{K}$ and $\mathcal{M}$ denote the set of users and UAVs, respectively. Here we follow a time-slotted synchronous communication framework \cite{sau2024drams} with slot duration $\delta_t$ and $N$ consecutive slots constitute a ``macro'' slot. In other words, the macro slot can be expressed as
\vspace{-2mm}
\begin{equation}
    N\delta_t=\delta_t^{(0)}+\delta_t^{(1)}+\ldots+\delta_t^{(N-1)}=\sum\limits_{i=0}^{N-1}\delta_t^{(i)}, 
\end{equation}
where $\delta_t^{(i)}$ corresponds to the $i$-th slot of a particular macro slot. Here, the user-UAV assignment is determined at the beginning of each slot, while the position of the UAVs change at the beginning of each macro slot. Accordingly, we introduce an indicator variable $i_{m,k}$ as follows.
\vspace{-2mm}
\begin{equation}\label{assgn_indicator}
i_{m,k} =
\begin{cases} 
1, & \text{if the UAV $m$ serves the user $k$,} \\
0, & \text{otherwise.}
\end{cases}
\end{equation}
Also, each UAV $m \in \mathcal{M}$ has a limitation of how many users it can serve, i.e., each UAV can serve at most $n_m$ users. At the beginning of an arbitrary macro slot, the position of the UAVs is $\chi_m = (x_{m}^{(0)}, y_{m}^{(0)}, z_{m}^{(0)}) : m \in \mathcal{M}$ and 
if $\phi_m$ is the UAV location in the previous macro slot, the distance traveled by the UAV $m \in \mathcal{M}$ with velocity $V_m$ is $d_m=||\phi_m-\chi_m||$ and the corresponding energy consumption is  \cite{zeng2019energy} 
\vspace{-1mm}
\begin{equation}\label{energy}
    E(V_m) \approx P_0 \left( 1 + \frac{3V_m^2}{U_{\text{tip}}^2} \right) + \frac{P_i v_0}{V_m} + \frac{1}{2} d_0 \rho s A V_m^3,
\end{equation}
where $P_0,P_i,U_{\text{tip}},v_0,d_o,s,A$, and $\rho$ are the system parameters as described in \cite{zeng2019energy}.

At the beginning of a macro slot, to provide enhanced service to the users, we move the UAVs in such a way that net system energy consumption is minimum and also each UAV needs to move to their new positions within a desired amount of time, say $t$, so that there is not much delay in establishing the communication. In this context, we construct a bipartite graph $\mathcal{G}_m$ with the two independent sets $\chi \coloneqq \{\chi_m : m \in \mathcal{M}\}$  and  $\phi \coloneqq \{\phi_j :  j \in \mathcal{M}\}$.  An edge from $\chi_m $ to $\phi_j$ exists iff the location is reachable by the UAV within time $t$ and the weight of the edge is defined as the total energy consumed by an UAV for traveling from $\phi_j$ to $\chi_m $ and this distance is denoted by $d_{m,j}$. Hence, we need to find the minimum weight matching for $\mathcal{G}_m$ and for this, we introduce an indicator variable
\vspace{-2mm}
\begin{equation}\label{UAV_assgn_indicator}
c_{\chi_m,\phi_j} =
\begin{cases} 
1, & \text{if UAV at $\phi_j$ moves to $\chi_m$,} \\
0, & \text{otherwise.}
\end{cases}
\end{equation}

\subsection{Throughput Calculation}
With respect to $\delta_t^{(i)}$ $\forall$ $i=0,\ldots,N-1$ of an arbitrary macro slot, the received signal at the UAV $m \in \mathcal{M}$ from the user $k \in \mathcal{K}$ is
\vspace{-2mm}
\begin{equation}
    s_{r,m,k}=\sqrt{P_{m,k}} g_{m,k} s_{t,m,k}+n_o,
\end{equation}
where $P_{m,k}$ is the path loss dependent received power, $g_{m,k}$ is the complex channel coefficient, $s_{t,m,k}$ is the
transmitted signal, and $n_o$ denotes the additive white Gaussian noise (AWGN) with power $N_0$. Here, $|g_{m,k}|$ is a Rician or Rayleigh random variable depending on whether it corresponds to the LoS or NLoS channel, respectively \cite{sau2025priority}. Moreover, $d_{m,k}$ is the Euclidean distance between UAV $m$ and user $k$ and $P_{m,k}$, which depends on the 73 GHz path loss model (PL) considered \cite{akdeniz2014millimeter}, is characterized in terms of $d_{m,k}$ as
\vspace{-2mm}
\begin{equation}
    P_{m,k} [dBm]= P_k + G_k+G_m-PL_{m,k}(d_{m,k}),
\end{equation}
where $P_k$ is the transmit power of user $k$, $G_k(G_m)$ is the gain of the transmit (receive) antenna, and
\vspace{-2mm}
\begin{equation}
    PL_{m,k}(d_{m,k})[dB] = \alpha + 10\beta\log_{10}(d_{m,k}).
\end{equation}
Here, $ \alpha $ and $ \beta $ are parameters of the LoS and NLoS models respectively. Therefore, the corresponding throughput for user $k$ at UAV $m$ is
\vspace{-2mm}
\begin{equation}\label{d_rate}
    r_{m,k}=B_w \log_2 \left( 1+\frac{P_{m,k}|g_{m,k}|^2}{N_0} \right),
\end{equation}
where $B_w$ is the channel bandwidth.



\subsection{User Mobility Model}
The initial positions of the users at the beginning of $\delta_t^{(i)}$ $\forall$ $i=0,\ldots,N-1$ are denoted by $\xi_{k,i}^{(0)} = (x_{k,i}^{(0)}\ , \ y_{k,i}^{(0)}\  , \ z_{k,i}^{(0)}) : k \in \mathcal{K}$. As we consider a mobile scenario here \cite{rwp}, the user $k \in \mathcal{K}$ moves with a velocity $v_{k,i} \in [0,V_k^{(max)}]$ and at the end of $\delta_t^{(i)}$, it can be anywhere inside the uncertainty circle $\zeta_{k,i}$ of radius $r_{k,i}=(v_{k,i}\cdot \delta_t)$ centered at $\xi_{k,i}^{(0)}$. We divide $\zeta_{k,i}$  into $n_{k} = 2\pi/\theta_k$ sectors of central angle $\theta_k$ each. Also, we consider a set $\Gamma_{k,i} = \{\gamma_{k,l}:1\leq l\leq q_{k,i}  \}$ of concentric circles centered at $\xi_{k,i}^{(0)}$, where $\gamma_{k,l}$ has radius $sl$ and $q_{k,i}=r_{k,i}/s$, s is our step size such that $s$ is a divisor of $r_{k,i}$ and $1\leq l \leq q_{k,i}$. Hence, the total $q_{k,i}n_k$ points are our possible search points, which are obtained by the intersection of the concentric circles and the sides of the sectors. Accordingly, the coordinates of these search points for the $k$-th user are $\xi_{k,i}^{(j)} = (x_{k,i}^{(j)},y_{k,i}^{(j)},z_{k,i}^{(j)})$ where $1\leq j\leq q_{k,i}n_k $. During the continuing slot, the user can be at any of these points with probability $p_{k,i} = 1/(q_{k,i}n_{k})$.

\subsection{LoS Detection Based on the Distribution of the Buildings} As demonstrated in \cite{li2022geometric}, the $x-y$ plane is divided into grids of equal size. Thereafter, the intersection points of these grids and projection of the line joining an UAV $m$ at $\chi_m$ and user $k$ at $ \xi_{k,i}^{(j)}$ denoted by $C_{m,k}^{(j)}$, is examined. If at any intersection point  $c_{m,k}^{(j)} \in C_{m,k}^{(j)}$, the building height $H_{c_{m,k}^{(j)}}$ is greater than the projected height $z_{c_{m,k}^{(j)}}$ then the link is NLoS. Accordingly we introduce an indicator variable $b_{\chi_m,\xi_{k,i}^{(j)}}$ denoting if there is a Los link between UAV $m$ and user $k$ as follows.
\vspace{-0.5mm}
\begin{equation}
b_{\chi_m,\xi_{k,i}^{(j)}} =
\begin{cases} 
1, & \text{if $ \forall\  c_{m,k}^{(j)} \in C_{m,k}^{(j)}\ :\ z_{c_{m,k}^{(j)}}>H_{c_{m,k}^{(j)}}$ ,} \\
0, & \text{if $\exists\  c_{m,k}^{(j)} \in C_{m,k}^{(j)}\ :\ z_{c_{m,k}^{(j)}}\leq H_{c_{m,k}^{(j)}}$ }
\end{cases}
\end{equation}

\subsection{User Mobility Based Throughput Modeling}
Suppose user $k$ is connected to UAV $m'$ and $m''$ during the slots $\delta_t^{(i)}$ and $\delta_t^{(i-1)}$, respectively. In case we have $m' \neq m''$, the UAV requires a finite amount of time $\epsilon$, at the beginning of $\delta_t^{(i)}$, to disconnect from $m'$ and connect with $m''$. This is why there is a possibility, that in spite of obtaining a higher throughput with the UAV $m'$, the total transferred data can be lesser than if user $k$ is kept assigned with UAV $m''$. Consequently, we introduce an indicator variable as follows.
\vspace{-2mm}
\begin{equation}\label{assgn2}
a_{m',m''}^{(k)} =
\begin{cases} 
0, & \text{if $m'=m''$,}\\ 
1, & \text{otherwise.}
\end{cases}
\end{equation}
In this work, we are considering a 73 GHz pathloss model, where the NLoS links are nearly infeasible due to high frequency. Therefore, we consider only the LoS links between the UAVs and the users. Since the user $k$ can be in any of $q_{k,i}n_{k}$ coordinates inside $\zeta_{k,i}$, that is, the uncertainty circle, by using \eqref{d_rate} and \eqref{assgn2}, we formulate an expected throughput definition. Accordingly, the throughput between UAV $m$ and user $k$ during slot $\delta_t^{(i)}$ depending on LoS or NLoS as
\vspace{-1mm}
\begin{align}
    R_{m,k}^{(j)}(\chi_m, \xi_{k,i}^{(j)}, b_{\chi_m,\xi_{k,i}^{(j)}})&=b_{\chi_m,\xi_{k,i}^{(j)}} r_{m,k}^{(j)},
\end{align}
where $r_{m,k}^{(j)}$ from \eqref{d_rate} is the throughput of user $k$ at $\xi_{k,i}^{(j)}$, which is associated with the UAV $m$ at $\chi_m$ during slot $\delta_t^{(i)}$. Hence, the total expected transferred data for user $k$, by considering all the search points inside uncertainty circle during current time interval, is formulated as
\vspace{-2mm}
\begin{equation}\label{thpt2}
    R_{m,k}(\chi_m)=\sum_j^{q_{k,i}n_{k}} (p_{k,i}\cdot R_{m,k}^{(j)}).
\end{equation}
Now, as discussed before, during the current slot $\delta_t^{(i)}$,  it may be advantageous if the user is still connected with the UAV, to which it was assigned during $\delta_t^{(i-1)}$. By using \eqref{assgn2} and \eqref{thpt2}, the overall transferred data during current time slot is
\vspace{-1mm}
\begin{equation}\label{thpt}
    R_{m,k}^{\text{(total)}} = (\delta_t - a_{m,m'}^{(k)}\cdot\epsilon)\cdot R_{m,k}.
\end{equation}

\subsection{Problem Formulation}
Here, our entire problem can be broken down into two distinct optimization problems: (1) Movement of UAVs in each macro slot, and (2) user-UAV assignment in each slot.
By using \eqref{energy} and \eqref{UAV_assgn_indicator}, the first optimization problem can be formulated as a minimum weighted matching problem $\rm (P1)$ for the bipartite graph $\mathcal{G}_m$, i.e.,
\vspace{-2mm}
\begin{equation}\label{bipartite_matching}
    {\rm (P1):} \qquad
    \min_{c_{\chi_m,\phi_j}}\;\sum_{j}\sum_{m}\frac{c_{\chi_m,\phi_j}E(V_j)\cdot d_{m,j}}{V_j}
\end{equation}
subject to
\begin{align*}
    \sum_{m} c_{\chi_m,\phi_j}= 1,\sum_{j} c_{\chi_m,\phi_j}= 1, \: \text{and} \: \sum_{m} \frac{c_{\chi_m,\phi_j}d_{m,j}}{V_j} \leq t.
\end{align*}
Next, the second optimization problem regarding the user-UAV assignment issue is formulated by using \eqref{assgn_indicator} and \eqref{thpt} as
\vspace{-2mm}
\begin{equation}\label{GAP}
   {\rm (P2):} \quad \max_{a_{m,m'}^{(k)},\ i_{m,k}} \, R_{total} = \sum_{k=1}^K \sum_{m=1}^Mi_{m,k} \, R_{m,k}^{\text{(total)}}(\chi_m)
\end{equation}
\begin{equation*}
\text{subject to} \: \: \sum_{m=1}^M i_{m,k}\leq 1 \:\: \text{and} \:\: \sum_{k=1}^K i_{m,k} \leq n_m,
\end{equation*}
for all $m \in \mathcal{M}$ and $k \in \mathcal{K}$, respectively. It is evident that \eqref{GAP} is a generalized assignment problem (GAP), which has been proven NP-hard in \cite{ozbakir2010bees}.

\section{Proposed Strategy}\label{prop_str}
In this section, we propose strategies for solving $\rm (P1)$ and $\rm (P2)$, as stated above. Note that the location of the UAVs change at the beginning of each macro slot, whereas the user-UAV assignment is decided in each slot.  Algorithm \ref{alg:hung_algo} finds the energy efficient movement of the UAVs. To do this, the users must first be clustered, and we propose the Algorithm \ref{alg:updated_kmeans} for this purpose. The user-UAV assignment for the remaining slots in the current macro slot without repositioning the UAVs is obtained by Algorithm \ref{alg:assignment}. We use the notion of `priority' for doing the user-UAV assignment, where user $k$ can wait for a maximum of time $t_k$. At the beginning of $\delta_t^{(i)}$, if the user $k$ has already waited for time $t_w$, its priority is defined as
\begin{equation}\label{priority}
    pr_k = \frac{t_w}{t_k}.
\end{equation}

\begin{algorithm}[h!]
\caption{Proposed Algorithm for optimal UAV movement}
\label{alg:updated_kmeans}
\begin{algorithmic}[1]

\State Initialize random locations $(x_m, y_m)$ for UAVs, set iteration number $n = 0$;
\State Set $pr_{k}:$ priority of user $k$ using \eqref{priority};

\State Sort $pr_k$'s in decreasing order and let $Pr$ be this order\;\label{step1};
\Repeat
    \State  $n = n + 1$, $A_m^{(n)} = \emptyset$;
    \State  $Free_{\text{UAV}} = \mathcal{M}$;
    \State  $i_{m,k}=0 ,\forall k \in \mathcal{K}, \forall m \in \mathcal{M}$;\label{step2}
    \For{each $pr_{k}$ in $Pr$}\label{step6}
        \If{$Free_{\text{UAV}}$ = $\emptyset$}
            \State Break;
        \EndIf   
        \State m = $\operatorname*{arg\,max}\limits_{m' \in \text{Free}_{\text{UAV}}}
         R_{k,m'}^{(total)}$;\label{step3}

        \If{$R_{m,k}^{(total)} > 0$}
        \State Set $i_{m,k}$ = 1;
        \State $A_m^{(n)} = A_m^{(n)} \cup \{k\}$;
        \EndIf
        \If{$|A_m^{(n)}|\geq n_m$}
        \State $Free_{\text{UAV}} = Free_{\text{UAV}}\setminus \{m\}$;
        \EndIf
    \EndFor \label{step7}
    \For{$m \in \mathcal{M}$}
        \State update $(x_m^{(n)} , y_m^{(n)} )$ using \eqref{subh1};\label{step4}
    \EndFor
    
\Until{convergence or maximum iterations is reached}
\For{$k \in \mathcal{K}$}
    \If{$i_{m,k} = 0 , \quad \forall m \in \mathcal{M}$}
    \State $pr_k = pr_k+\frac{\delta_t}{t_k}$;\label{step5}

    \EndIf
\EndFor    \\
\Return   $i_{m,k}$ ,  $\chi_m^{(n)} =(x_m^{(n)} , y_m^{(n)}, z_m^{(0)} )$; 
\end{algorithmic}
\end{algorithm}

\subsection{Initial Clustering of the UAVs after Each Macro Slot}
Initially we use a modified version of K-Means  clustering algorithm \cite{bottou1994convergence} to determine the balanced user cluster and the UAV positions. Conventional K-Means clustering is done by using the distance between the users and the clusters centroids. However, as the LoS plays a crucial role in  mmWave communications, we use throughput as the metric to cluster the users.
Note that, always assigning the users to their best choices can make the clusters highly imbalanced and hence, to mitigate this problem, we consider the user's priorities as defined in \eqref{priority}. Let $A^{(n)}_m$ be the set of all users  assigned  to UAV $m$ during the $n$-th iteration. At each iteration of the proposed Algorithm 1, we iteratively  select the user with the maximum priority and assign it to the UAV which has not exceeded its capacity and gives the user its best non-zero throughput. If no such UAV is found, the user will remain unattended and during each iteration, we update the 2D co-ordinates of the UAVs as
\vspace{-2mm}
\begin{equation}  \label{subh1}
    (x_m^{(n)} , y_m^{(n)} ) = \frac{\sum_{k \in A_m^{(n)}} R^{(total)}_{m,k}*(x_{k,0}^{(0)} ,  y_{k,0}^{(0)})}{\sum_{k \in A_m^{(n)}}R^{(total)}_{m,k}},
\end{equation}
where $R^{(total)}_{m,k}$ is calculated by \eqref{thpt}. The process will continue until the positions of the UAVs do not change or maximum number of iteration is reached. After the algorithm stops, the priority of each user is updated, such that all the users have a fair chance of getting served in the upcoming slots.

\begin{algorithm}[h!]
\caption{Energy efficient movement of UAVs}
\label{alg:hung_algo}
\begin{algorithmic}[1]
\State Get the new cluster centroids $\{\chi_m : m\in \mathcal{M}\}$ using \textbf{Algorithm \ref{alg:updated_kmeans}};\\
\textbf{Define:} $\{\phi_j  : j\in \mathcal{M}\}$, locations of UAV in the previous macro slot;\\
\textbf{Define:} $e_{i,j}$ $\triangleq$ energy consumption for the UAV $j$ movement from $\phi_j$ to $\chi_i$ using equation \eqref{energy}.;
\State Construct the cost matrix $C_{m\times m} = (c_{i,j})$;
\If{$\chi_i$ is reachable from $\phi_j$ by UAV $j$ within time $t$}
\State $c_{i,j} = e_{i,j}$;
\Else 
\State $c_{i,j} = \infty$;
\EndIf
\State Solve assignment problem (P1) using \textbf{Hungarian} method;\\
\Return   Assignment matrix and total energy consumption;
\end{algorithmic}
\end{algorithm}

\subsection{Energy Efficient Movement of the UAVs}
{After the clustering of the users is done by using Algorithm 1, we get the new assigned centroids for the $m$ UAVs. Now, as we need to solve (P1), we employ the well-known Hungarian method \cite{kuhn1955hungarian}. For assigning $m$ jobs (M cluster centroids) to $m$ machines (M UAVs), time complexity for the Hungarian method is $O(M^3)$\cite{kuhn1955hungarian}. Let $C_{M\times M}$ denote the $M\times M$ cost matrix, where the $(i,j)$-th entry $c_{i,j}$ is the energy consumption for moving the UAV $j$ from $\phi_j$ to $\chi_i$. If the time taken for this movement is higher than $t$, $c_{i,j}$ will be considered as $\infty$. Finally, by solving (P1) using Algorithm \ref{alg:hung_algo}, we get the assignment matrix for optimal UAV movement and the total energy consumption for the same.

\subsection{Capacity-constrained user-UAV Assignment}
During the entire macro slot, the position of the UAVs remain unchanged. As the GAP \eqref{GAP} is NP-hard, we propose a greedy algorithm (Algorithm \ref{alg:assignment}), which takes care of both the capacity-constrained scenario as well as the aspect of throughput maximization. Initially, by using \eqref{thpt}, each user is assigned to the UAV which gives it best throughput, hence resulting in high imbalance. To take care of this problem, we define three sets, namely, $Full$, $Free$, and $Assign$ for keeping track of the limit-exceeded UAVs, the UAVs with available slots, and the unassigned users, respectively. We iteratively select the UAV $m \in Full$ and retain the top $n_{m}$ users based on their priority and the remaining users are included in $Assign$ for further assignment. For each user $k \in Assign$ and each UAV $m \in Free$, we first calculate its \emph {sacrifice value} $S_{m,k}$, defined as the reduction in throughput for not being allocated to its best choice, say UAV $m_k$. That is, $S_{m,k}=R_{m_k,k}^{\text{(total)}}-R_{m,k}^{\text{(total)}}$. We select the user $k \in Assign$ with the least sacrifice, assign it to the corresponding UAV, and remove it from $Assign$. If in this process, some UAV reaches its limit, we remove it from $Free$ and repeat the process with remaining users in $Assign$ and remaining UAVs in $Free$. Accordingly, we update the priorities of the users.

\vspace{-2mm}

\begin{algorithm}[h!]
\caption{User-UAV assignment}
\label{alg:assignment}
\begin{algorithmic}[1]
\State \textbf{Define} $A_m$ $\triangleq$ the set of users assigned to UAV $m$;
\For{$k \in \mathcal{K}$}
    \State $m_k = argmax_{m \in \mathcal{M}}R_{m,k}^{(total)}$;
    \State $A_{m_k} = A_{m_k} \cup \{k\}$;
    
\EndFor
\State $Full = \{m :|A_m|\geq n_m\}$ ,  $Free = \{m :|A_m|< n_m\}$;
\State $Assign = \emptyset$;
\For{$ m \in Full $}
    \State Sort $A_m$ by decreasing order of $pr_k, \ \forall k \in A_m$;
\nonumber
    \State Let $X_m$ be the top $n_m$ users in the sorted order;
    \State  $Assign = Assign\cup(A_m\setminus X_m)$;
    \State Set $i_{m,k} = 1$ $\forall$ $k \in X_{m}$ ;
\EndFor
\State $S_{m,k}=R_{m_k,k}^{\text{(total)}}-R_{m,k}^{\text{(total)}}\; \forall k  \in Assign, \forall m \in Free  $;
\While{Assign $\neq \emptyset$}
        \State $(m^*,k^*) = \operatorname*{arg\,min}\limits_{m \in \text{Free},\;k \in \text{Assign}}  S_{m,k}$;

        \If{$R_{m^*,k^*}^{(total)}= 0$}
            \State $i_{m,k*}=0$ \quad $\forall m \in \mathcal{M}$;
        \Else
            \State Set $i_{m^*,k^*}=1$;
            \State $A_{m^*} = A_{m^*} \cup \{ k^* \}$;
        \EndIf  
    
    \State $Assign = Assign\setminus \{k^*\}$;
    \If{$|A_{m^*}|\geq n_m$}
        \State $Free = Free\setminus\{m^*\}$;
    \EndIf    
        
\EndWhile

\For{$k \in \mathcal{K}$}
    \If{$i_{m,k} = 0 , \quad \forall m \in \mathcal{M}$}
    \State $pr_k = pr_k+\frac{\delta_t}{t_k}$;
    \EndIf
\EndFor\\
\Return $i_{m,k}\;\forall m\in \mathcal{M},\;\forall k \in \mathcal{K}$;
\end{algorithmic}
\end{algorithm}

\subsection{Discussion on the Computational Complexity}
\begin{enumerate}
    \item \textbf{Algorithm 1:} Lines $2$ and $3$ requires $O(K)$ and $O(K\log(K))$ respectively. Each iteration of the loop from lines $4$-$24$ require $O(KM)$. If $I$ denotes the maximum iteration, then lines $4$-$24$ require $O(IKM)$. Lines $25$-$29$ requires $O(KM)$. Hence the total complexity of this algorithm is $O(IKM+K\log(K))$.

    \item \textbf{Algorithm 2:} Algorithm \ref{alg:hung_algo}  calls Algotrithm \ref{alg:updated_kmeans} to obtain the initial clustering  which has complexity $O(IKM+K\log(K))$ and subsequently uses Hungarian method for solving (P1), which has complexity $O(M^3)$. Hence, the total complexity is $O(IKM+K\log(K))+O(M^3)$.
    
    \item \textbf{Algorithm 3:} Lines $2$-$5$ and $6$ require $O(KM)$ and $O(M)$ respectively. Lines $8$-$13$ require $O(KM)$. Line $14$ requires $O(KM)$. Lines $15$-$27$ and $28$-$31$ require $O(K^2M)$ and $O(KM)$ respectively. Hence the total complexity of this algorithm is $O(K^2M)$.

\end{enumerate}


\begin{figure*}[t]
 \begin{subfigure}[b]{.33\textwidth}
    \centering
    \includegraphics[width=0.98\linewidth]{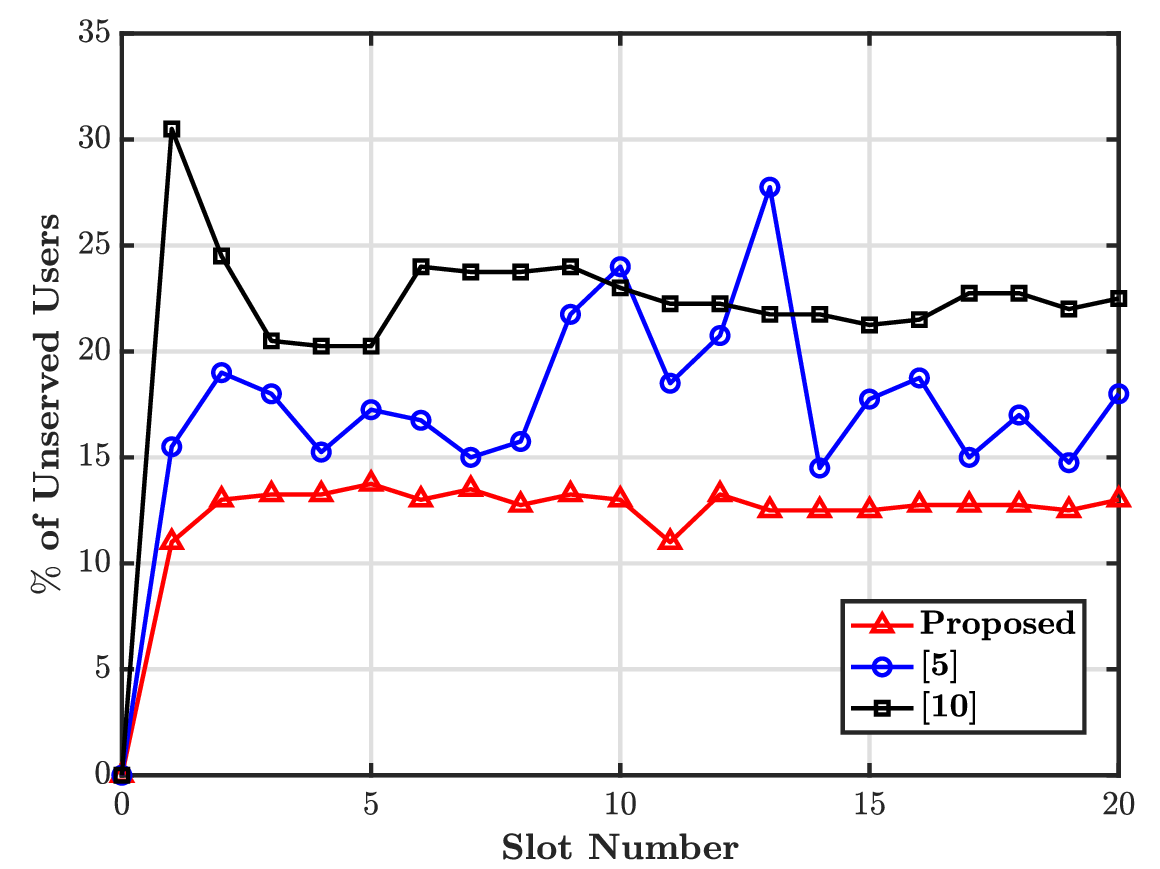}
    \vspace{-2mm}
    \caption{}
    \vspace{-2mm}
    \label{unserved}
\end{subfigure}
\begin{subfigure}[b]{.33\textwidth}
    \centering
    \includegraphics[width=0.98\linewidth]{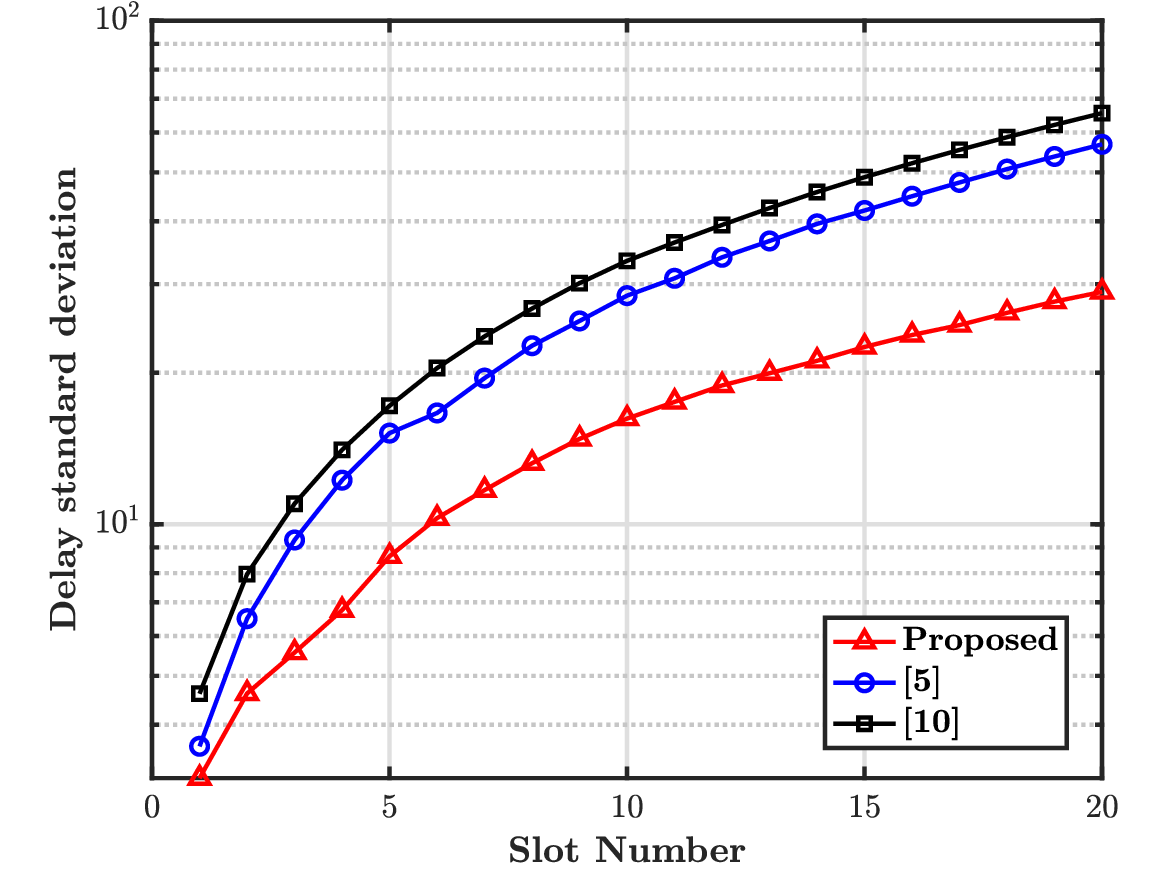}
    \vspace{-2mm}
    \caption{}
    \vspace{-2mm}
    \label{delay}
\end{subfigure}
\begin{subfigure}[b]{.33\textwidth}
    \centering
    \includegraphics[width=0.98\linewidth]{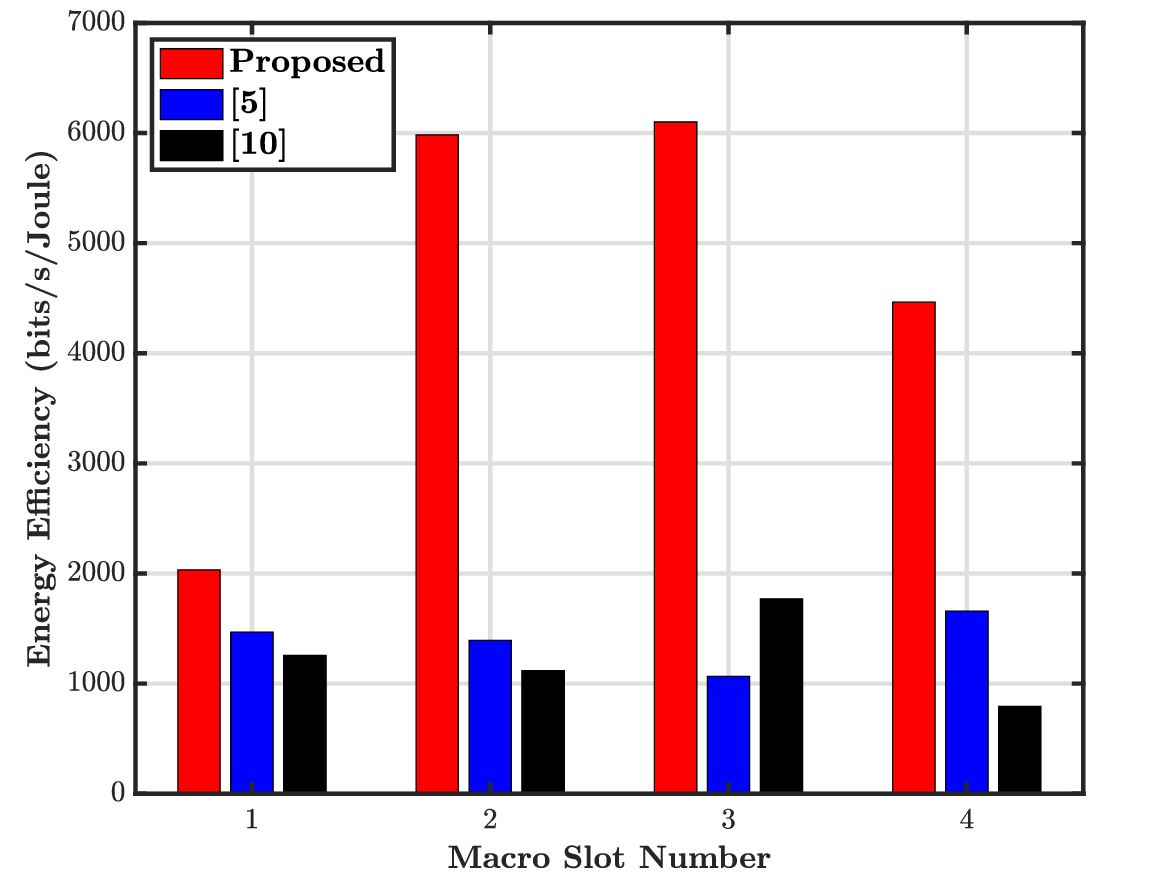}
    \vspace{-2mm}
     \caption{}
    \vspace{-2mm}
    \label{ee}
\end{subfigure}
\caption{\footnotesize  Impact on (a) $\%$ of unserved users, (b) standard deviation of the user delay, and (c) system energy efficiency.}
\vspace{-6mm}
\end{figure*}

\begin{table}[h!]
    \centering
      \caption{\small Simulation Parameters.} 
   \vspace{-2mm}
    \label{tab:sample}
    \resizebox{0.8\columnwidth}{!}{%
    \begin{tabular}{|c|c|}
        \hline
        Parameters & Values \\
        \hline
        Communication region   & 300m$\times$300m  \\
        Number of Users   & $|\mathcal{K}|=400$  \\
        Number of UAVs & $|\mathcal{M}|=6$\\
        Number of orthogonal frequencies & $n_m = 62$\\
        Altitude of User receivers & 1.5m\cite{li2022geometric}\\
        Flight altitude of the UAVs & [22m,150m]\cite{li2022geometric}\\
        Carrier Frequency & 73GHz\cite{akdeniz2014millimeter}\\
        Parameters for LoS links & $\alpha = 69.8 , \beta = 2$\cite{akdeniz2014millimeter}\\
        
        \hline
    \end{tabular}
    }
    \vspace{-2mm}
\end{table}

\section{Numerical Results}\label{result}
In this section, we discuss the performance of the proposed framework and also compare it with respect to the existing benchmark schemes. While the simulation parameters are given in Table \ref{tab:sample}, the performance is evaluated in terms of the percentage of unserved users, the standard deviation (SD) of the user delay, and the energy efficiency of the system. Note that, for the purpose of generation of results, we have considered $10$ time slots in a single macro slot, a transmission power of $30$ dBm, and a Rician fading scenario with $K=2$.

Fig. \ref{unserved} investigates the impact of our proposed framework on the percentage of the unserved users for each slot. It is interesting to observe, that unlike \cite{li2022geometric} and \cite{malinen2014balanced}, the proposed algorithm results in a significantly smaller percentage of unserved users. Note that assigning users to their best choice in terms of pathloss is not always the optimal solution, which results in more users being unserved. Moreover, nearest UAV assignment in scenarios of non-existent LoS is also not the desired solution. Hence, as our proposed algorithm takes these realistic factors into consideration, this results in an enhanced system performance.


Further, Fig. \ref{delay} illustrates the impact of our strategy on the SD of the user delay. Specifically, this metric emphasizes on the aspect of whether all the users in the system are getting access to the UAVs and not only a handful of them, where it is desirable to have a smaller SD, i.e., the lesser, the better. In this context, we observe that, for all the schemes, SD monotonically increases with the slot number. However, for any fixed slot number, our framework results in a significantly lower SD with respect to the other benchmark schemes. The reason for this is that, as explained earlier, our scheme results in a smaller $\%$ of unserved users and also, in each slot, the previously unserved users are always getting a fair chance of accessing a UAV for their communication purpose.


Finally, Fig. \ref{ee} demonstrates the impact of our scheme on the system energy efficiency, which is defined as the ratio of total throughput and total energy consumption. We observe that our framework significantly outperforms the existing schemes. This is based on the use of a Hungarian algorithm-based framework along with the fact that we are taking into account the existence/absence of LoS user-UAV links and not only the pathloss as the absolute measure for the UAV assignment problem. Moreover, by considering the user mobility-based scenario, we avoid frequent UAV movement in each slot but only once in each macro slot. Furthermore, after each macro slot, we provide a set of locations, where the UAVs are needed to be moved. Note that, we do not emphasize on having a particular UAV at any particular location. But, by considering minimum UAV energy consumption and the location of the UAVs in the previous macro slot, we decide which UAV needs to be relocated to which of the suggested locations. This also enhances the system energy efficiency, as we guarantee a minimum UAV displacement scenario.



\section{Conclusion}\label{con}
In this work, we have investigated the aspect of energy-efficient movement of UAVs and capacity-constrained user-UAV assignment in a multi-user multi-UAV framework. By considering the priority of the users and the importance of having an LoS for the user-UAV assignment, the proposed framework results in reduced per user delay and enhanced system energy efficiency. We discussed the complexity of the proposed algorithms and the numerical results demonstrated the enhanced performance of our framework with respect to the existing benchmark schemes. 

\bibliography{refers}
\end{document}